\documentclass[manuscript]{aastex}

\newcommand{\integral}{{\it INTEGRAL} }

\def\source{IGR~J16328-4726}
\def\be{\begin{equation}}
\def\ee{\end{equation}}

\shorttitle{\source\/}
\shortauthors{Fiocchi et al.}
\begin{document}
\title{\source\/: a new candidate Supergiant Fast X-ray Transient}
\author{M. Fiocchi\altaffilmark{1}, V. Sguera\altaffilmark{2}, A. Bazzano\altaffilmark{1}, L. Bassani\altaffilmark{2}, A. J. Bird\altaffilmark{3}, L. Natalucci\altaffilmark{1}, P. Ubertini \altaffilmark{1}}
\altaffiltext{1}{Instituto di Astrofisica Spaziale e Fisica Cosmica di Roma (INAF). Via Fosso del Cavaliere 100, Roma, I-00133, Italy}
{\altaffiltext{2}{Instituto di Astrofisica Spaziale e Fisica Cosmica di Bologna (INAF). Via Gobetti 101, Bologna, I-40129, Italy}
\altaffiltext{3}{School of Physics and Astronomy, University of Southampton, Highfield, SO17 1BJ, UK}
\begin{abstract}
The unidentified  source \source\/ was covered with \integral\/ observations for a long period ($\sim$9.8Ms) and was undetectable for most of the time while showing a very recurrent micro-activity with a duration from tens minutes to several hours.
We report the discovery of two strong outbursts started at 53420.65 MJD  and 54859.99 MJD respectively,
the first  with a duration of  $\sim$1 hour and the second with a lower limit on the duration of $\sim$3.5 hours. Furthermore, the source have been detected in nine other short pointings with significance between 4 and 5 $\sigma\/$ as well as in a one of the revolution (during the exposure $\sim$ 130 ks) at a significance level of $\sim$7 $\sigma\/$.
The stronger outburst spectrum is well described by a power law model
with a photon index of $\sim$2.0 and a flux of $\sim$3.3$\times~10^{-10}~erg~cm^{-2}~s^{-1}$ in the 20--50~keV energy band.
The weaker outburst and revolution spectra show the same spectral shape and different fluxes.
The combined timing and spectral properties observed during the outburst,  the recurrent nature of this transient source, the Galactic Plane location, a dynamic range $>$170 in the 0.3-10 keV band and $>$165 in the 20-50 keV and the presence of a IR star in the error circle of an XRT/Swift pointing are suggesting this source as a member of the class of the Supergiant Fast X--ray Transients.
\end{abstract}




\section{Introduction}
IGR J16328-4726 is a poorly studied X-ray source first reported by Bodaghee et al. (2007) as a transient unidentified source, located in the Galactic Plane. It is listed as a "blended" variable source in the 4th IBIS/ISGRI Catalog (Bird et al. 2010). The blending is due to the vicinity of the bright transient X-ray binary 4U~1630-47 with a 20-100 keV flux of $4.41\times10^{-11}erg~cm~^{-2}s^{-1}$.
This \source\/ is present in the Swift BAT 39 month all-sky survey (Cusumano et al. 2010) as a gamma source with a flux of $(4.8\pm0.3)\times~10^{-11}~erg~cm^{-2}~s^{-1}$ in the 14-70 keV energy band.
Grupe et al. (2009) reported the detection of a flare observed with Swift/BAT (at 07:54:27 UT on 2009 June 10)
which was followed up by a Swift/XRT observation, $\sim$385 s after the trigger.
Using 2754 s of XRT Photon Counting mode data, they determine the best source position yet at RA=248.1579 and Dec=-47.3951 with an uncertainty of 1.7 arcsec (90\% confidence). The spectrum was modeled with an absorbed power law having $\Gamma = 0.56^{+0.75}_{-0.68}$
and an absorption column density $N_H=8.1^{+5.7}_{-4.9}\times10^{22}~cm^{-2}$ in excess to the galactic value of $\sim~1.5\times10^{22}~cm^{-2}$. 
Recently Corbet et al. (2010)  have analyzed the Swift BAT 58 month survey (Baumgartner et al. 2010) light curve (from 2004-12-20 to 2009-09-30) in the 15-100 keV energy band and they report a highly significant modulation at a period close to 10 days.
This periodicity was interpreted as the orbital period of a high-mass X-ray binary, powered by accretion from the wind of a supergiant
companion. 

During the repeated Galactic Plane scanning performed with \integral\/ a new class of hard X-ray transient emitters have been discovered: the  Supergiant Fast X-ray Transients (SFXTs).
These systems display short outbursts with very fast rise time ($\sim$ tens of minutes) and typical duration of a few hours. The dynamical range observed in the 20-60 keV between quiescent ($L_X\sim10^{32}erg/s$) and peak ($10^{36}-10^{37} erg/s$) luminosities is quite high ($10^2-10^5$) while the X-ray spectra measured are often hard, like those characterizing HMXB hosting an X-ray pulsar.

Here we provide a detailed analysis of the timing and spectral properties of \source\/, showing that it is a highly variable source and a new candidate for the SFXT class.

\section{Observations and Results}
\subsection{INTEGRAL/IBIS results}

In order to provide new information on  \source\/, we have analyzed the INTEGRAL/IBIS-JEM-X
public database. We also reanalyzed XRT observations performed during a source flare.

The \integral\/ (Winkler et al. 2003) observations
are divided into uninterrupted $\sim$2000~s intervals, known as science windows (SCWs).
Data are extracted from each individual SCW and
processed using the Off-line Scientific Analysis
(OSA v9.0) software released by the
\integral\/ Scientific Data Centre (Courvoisier et al. 2003).
We analyzed all public IBIS/ISGRI (Ubertini et al. 2003) observations within
15$^{\circ}$ of the source, i.e. 3783 pointings performed between 2003-01-29 and 2009-03-20,
yielding  a total exposure time of about 9.8~Ms.
These runs were performed with AVES cluster, designed to optimize performances and disk storage
for the INTEGRAL data analysis by Federici et al. (2010).
Images, spectra and light curves were extracted for each Science Window in the 20-50 keV energy band.

\subsection {The image analysis}

For most of the observation time \source\/  was below the instrument sensitivity ($\sim$0.5 mCrab for an exposure time of $\sim$9Ms). 
It was detected in several single SCWs, in groups of the SCWs and also in a revolution without a clear periodicity, although this study is limited by the source coverage and by the
fact that \source\/ is only 0.2 degrees from the bright transient Black Hole 4U~1630-472 which often contaminates
the source emission, the IBIS/ISGRI angular resolution being of the same order of the distance between the two objects.
For this reason, we do not consider the activity periods associated with \source\/ during those times when 4U~1630-472 is detected with $SNR>4\sigma$.

\begin{figure*}[t]
\includegraphics[angle=0, width=0.7\linewidth]{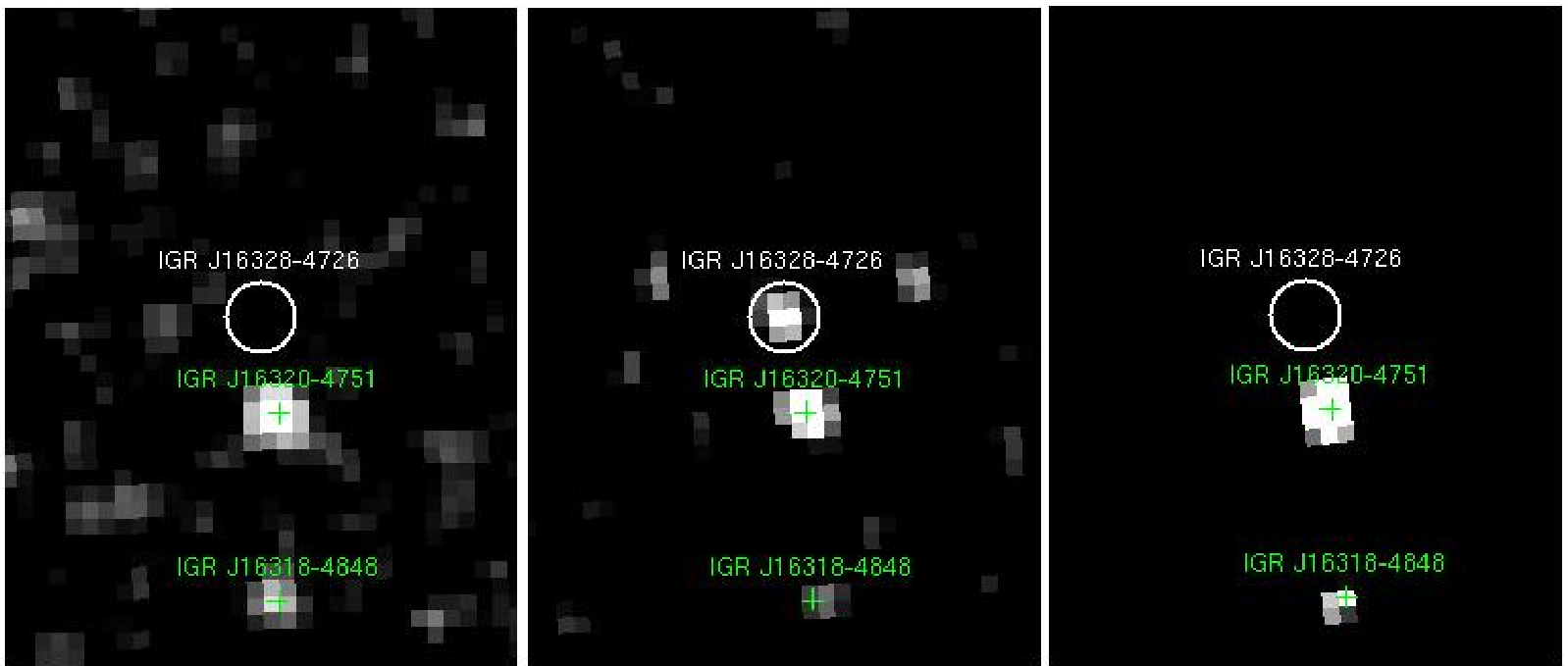}
\includegraphics[angle=0, width=0.7\linewidth]{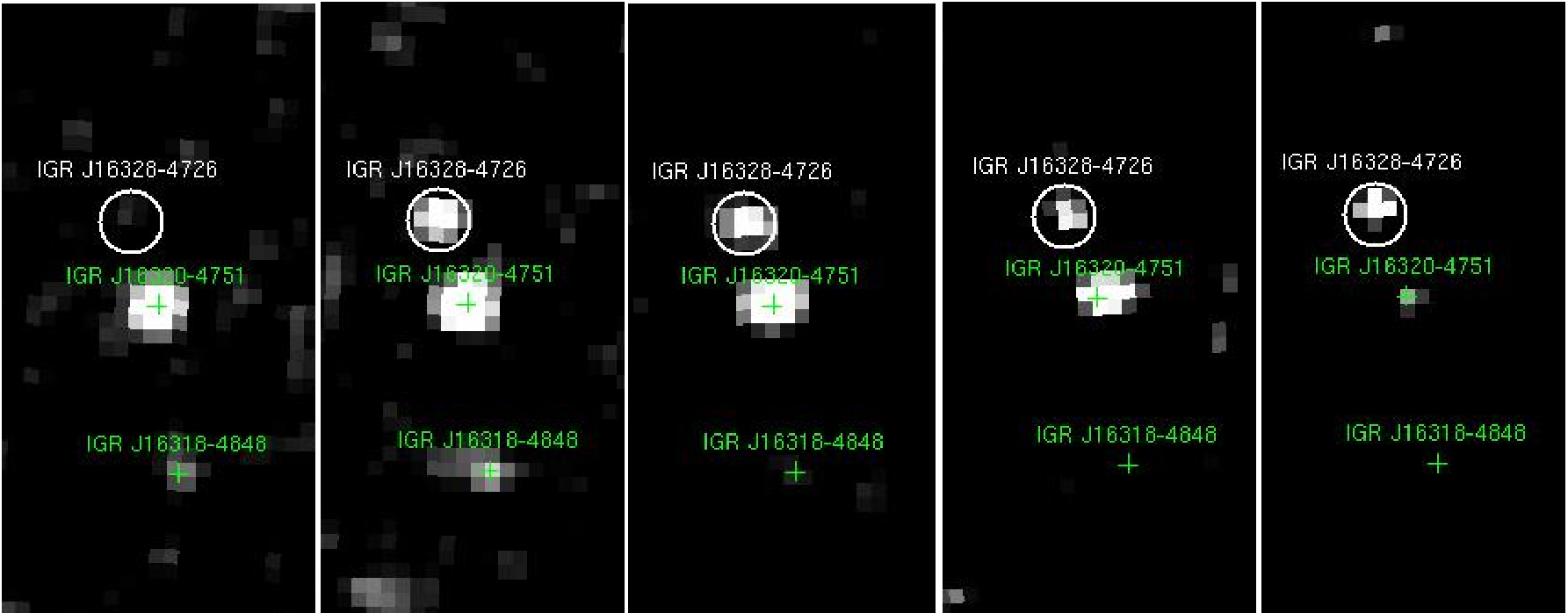}
\caption{Top panel: ISGRI Science Window image sequence (20--50 keV) of the first outburst (No. 1 in Table 1) of \source\/ (encircled).
The duration of each ScW is $\sim$ 2000 s. 
Significance of the detection, left to right, was $<$2$\sigma$,  6$\sigma$ and $<$2$\sigma$.
Bottom panel: ISGRI Science Window image sequence (20--50 keV) of the second outburst (No. 2 in Table 1) of \source\/ (encircled).
Significance of the detection, left to right, was $<$2$\sigma$,  11$\sigma$, 8$\sigma$, 4$\sigma$ and 7$\sigma$.}
\label{fig1}
\end{figure*}

We created an image for each SCW in the 20--50 keV energy band and searched for the detections of \source\/ above a threshold of 4 $\sigma$; we visually inspected all SCW images with a positive detection.
This SCW image inspection was very important because it allowed to further check on the presence of the near sources and so properly associate the hard X-ray emission of \source\/ compared to the one from the Black Hole 4U~1630-472.
This analysis has shown that \source\/ was active in 27 SCWs with a significance $>$4$\sigma$, but only 14 detections can be assigned to \source\/
for certain. In the remaining cases the near Black Hole 4U~1630-47 is bright in the image and detected with a signal to noise ratio higher than 4.

From the image analysis two strong
outbursts have been identified: the first in one SCW of revolution 287 and the second  in four consecutive SCWs of  revolution 768,
occurring on 53420.65 MJD  and 54859.99 MJD, respectively (see Table~\ref{tab:det}).
We assumed the beginning of the first SCW during which the source was detected as the start time
of the outburst and similarly the outburst stop time at the end of
the last SCW during which the source was detected.
Due to the sparse sampling of the
observations, we cannot determine precisely the duration
of these outbursts.
The first outburst lasted for $\sim$ 1 hour (see detail in Sec. 2.3), while for the second outburst the lower limit on the duration is $\sim$3.5 hours .
Figure 1 (top panels) shows the image sequence of three consecutive SCWs (20--50~keV) during which the first outburst was observed and in 
Figure 1 (bottom panels) we report the image sequence of five consecutive SCWs (20--50~keV) of the second one.
No data are available in the \integral\/ archive to study the outburst behavior immediately after.
Furthermore, \source\/ was also detected in 9 single SWCs (2-3 ks each)  and in one of revolution (using its whole duration) with a significance 4-5 $\sigma$ and 7 $\sigma$, respectively.

\begin{figure*}[t]
\includegraphics[angle=270, width=0.8\linewidth]{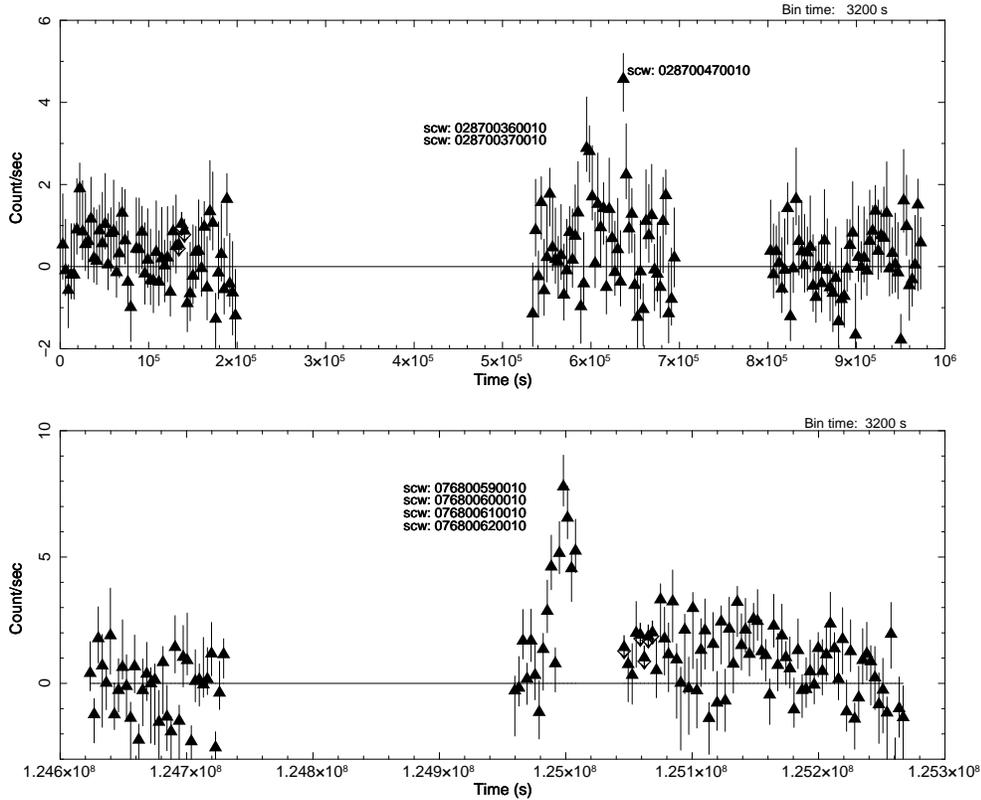}
\caption{ISGRI light curve (20--50~keV) of the first  (top)  and the second (bottom) outbursts of the \source\/ (see Table 1).
Time is in second since JD=2453413 19:49:01.380.}
\end{figure*}

An image mosaic using data from five SCWs of two strong outbursts has been created.
The scientific analysis of this mosaic shows
 \source\/  was detected with a significance level of $\sim$20$\sigma$ at RA(J2000)=248.150 and DEC(J2000)=-47.388,
with an error circle of 1.5 arcmin (at 90\% confidence level) and a flux of
$\sim$3.2$\times$10$^{-10}$~erg~cm$^{-2}$~s$^{-1}$ in the 20--50~keV energy band.
The position, flux and significance are slightly different from values reported by Bird at al. 2010.
We expect that this analysis, tuned to the specific case of \/source\/, should yield a better position estimate than the standard analysis performed for the survey catalog production.

\begin{figure*}
\includegraphics[angle=-90, width=0.5\linewidth]{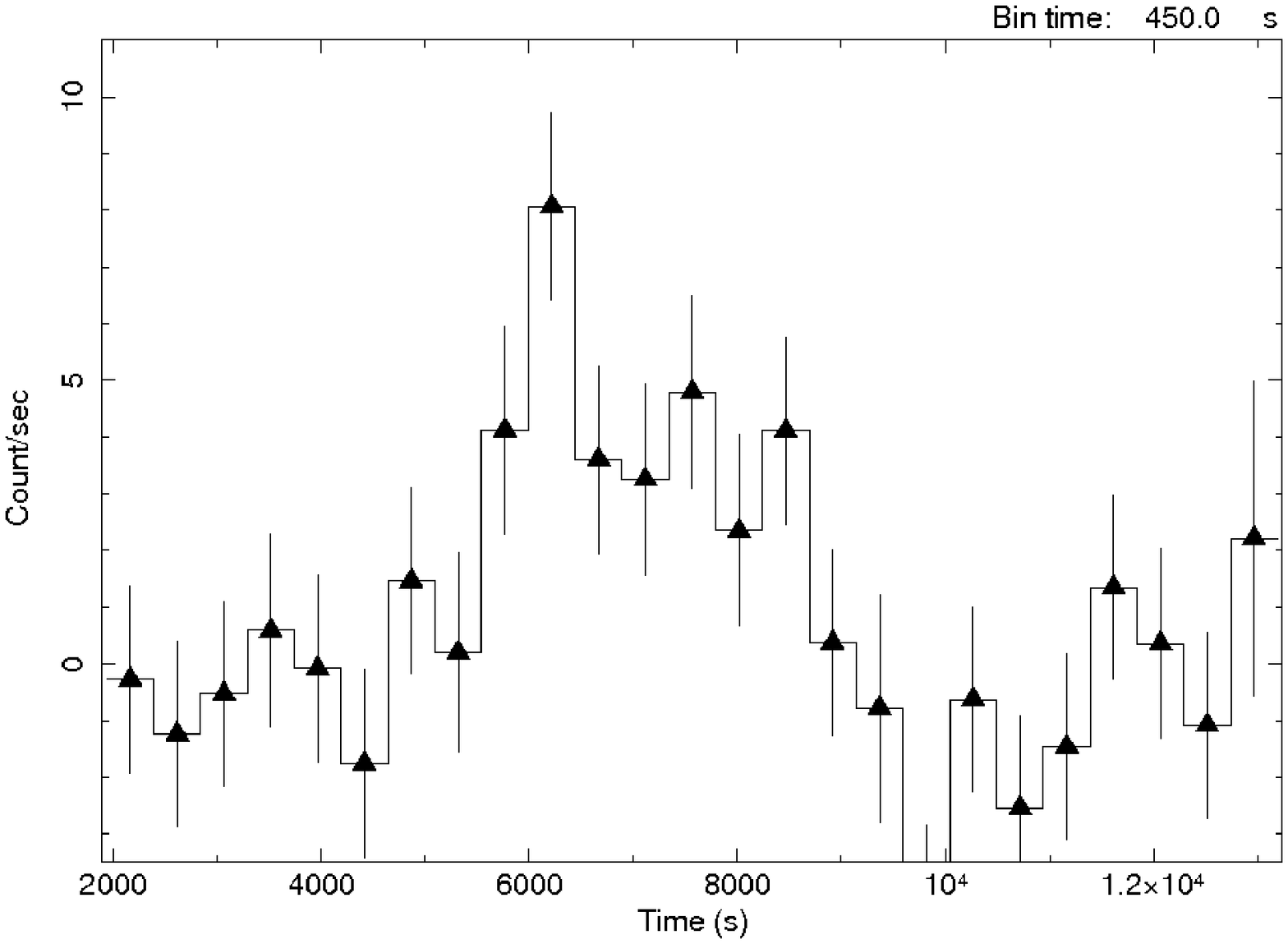}
\includegraphics[angle=-90, width=0.5\linewidth]{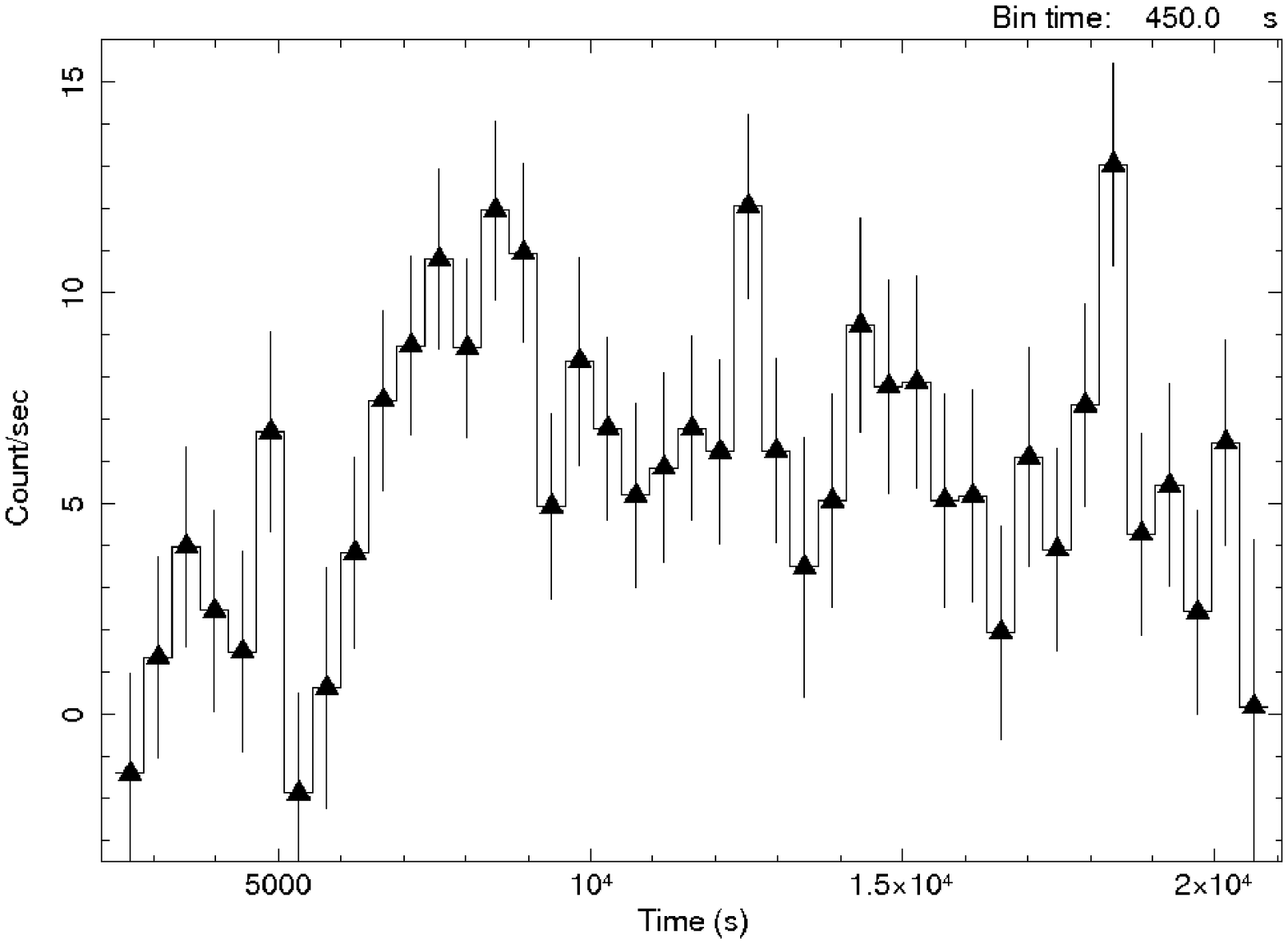}
\caption{Zoom of the ISGRI light curve (20--50~keV) of the first (left) and second (right) outbursts of the \source\/ (see Table 1).
Time is in seconds since JD=2453421 05:36:06.910 and 2454860 10:43:46.186 respectively for the first and second outbursts.}
\end{figure*}

\subsection {The light curve analysis}

ISGRI light curves (20--50~keV) are shown in Figure 2 for the first (top) and second (bottom) outbursts.
The observed outbursts have complex structures characterized by several fast flares with both rise and decay times less than one hour or lasting a few tens of minutes.
In the top panel we show the time behavior around the first outburst: \source\/ was below the IBIS detection limit for most of the time and becomes active
in two consecutive SCW
(028700360010 and 028700370010) and then in the SCW  028700470010 i.e. later 20 ks.
This last detection is the strong outburst of the revolution 287 that we are here describing as first outburts.
In the bottom panel the time behavior around the second outburst was shown: \source\/ was consistent with no emission for long time and
then active in the consecutive SCWs 076800590010, 076800600010, 076800610010 and 076800620010.
Figures 3 shows the light curve zoom at the time of the outbursts.
In the left panel is reported the first outburst behavior: \source\/ flux is initially consistent with zero, then suddenly turns on at
2005-02-19 15:32:46.91 UTC and quickly reaches the peak after
$\sim$ 15 minutes (15:47:46.91 UTC). Then
it drops to a lower flux level for $\sim$ 37 minutes and at 15:55:16.91 UTC the source completely disappears below the IBIS detection limit.\\
Right panel shows the second outburst behavior: \source\/ initially has a flux consistent with no emission, then suddenly turns on at 2009-01-29 00:30:36.00 and reaches the peak after $\sim$ 30 minutes and it remains active for more than three hours, after which unluckily there is not IBIS data coverage available in the \integral\/ archive.

\subsection {The spectrum analysis}

We extracted the ISGRI spectra during the two strong outburst periods and during the rev 351.
The spectrum observed during the first outburst can be fitted with
a single power law ($\chi^2/$d.o.f.=$4.1/3$) having a photon index
of 2.2$\pm0.9$ and a 20--50~keV
flux of $\sim$1.9$\times$10$^{-10}$~erg~cm$^{-2}$~s$^{-1}$.
The spectrum observed during the second and strongest outburst can be fitted with
a single power law ($\chi^2/$d.o.f.=$2.5/4$ ) having a similar photon index
(2.0$\pm0.4$) and a higher 20--50~keV
flux  ($\sim$3.3$\times$10$^{-10}$~erg~cm$^{-2}$~s$^{-1}$).
During revolution 351 the spectrum was well fitted with a simple power law model ($\chi^2/$d.o.f.=$4.2/3$),
having the same spectral index (2.6$\pm1.2$) and a flux of $\sim$4.4$\times$10$^{-11}$~erg~cm$^{-2}$~s$^{-1}$ in the 20-50 keV band.

\begin{figure}[t]
\includegraphics[angle=-90, width=0.7\linewidth]{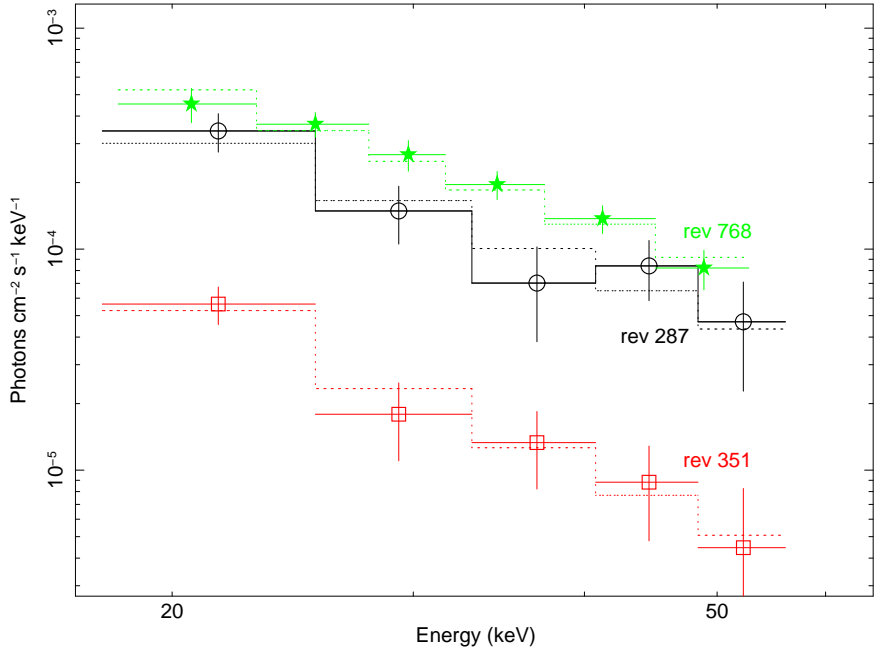}
\caption{ \source\/ photon spectrum in the 20--70~keV energy band from the first (rev. 287, circle), second (rev. 768, stars) outbursts and from revolution 351 (squares), fitted with a single power law model. See the text for the best fit parameters.}
\end{figure}

These spectra are shown in Figure 4.

As is typical for these HMXB systems, we also fitted these spectra with a thermal bremstralung model; the temperature are $43_{-24}^{+156}$ keV, $30_{-18}^{+170}$ keV and $51_{-19}^{+58}$ keV for the outbursts of revolution 287, 351 and 768, respectively.

The emerging picture is that this source shows strong variability on different time scales: two strong outbursts detected in single SCW or in a group of few SCWs, a weak detection for a long time (73 SCWs of rev 351) and some very weak detections lasting only one SCW.
The spectral analysis showed that the spectral shape is constant while the flux change by a factor up to $\sim$10.

In order to study the quiescent emission in the hard X-ray band, we obtained a mosaic image with exposure of $\sim$7 Ms using all available SCWs except for those in which both \source\/ and 4U~1630-47 were detected.
This analysis allows us to get a 3$\sigma$ upper limit for the flux of $2.5\times10^{-12}erg~cm^{-2}~s^{-1}$, giving a ratio between the outburst and quiescent fluxes $>165$ in the 20-50 keV energy band.

\subsection{INTEGRAL/JEM-X results}

We analyzed all public JEM-X pointings within
4.5$^{\circ}$ of the source, performed between 2003-01-29 and 2009-03-20,
yielding  a total exposure time of  372 ks.
Unfortunately, bacause of the relatively small field of view of JEM-X, there is no JEM-X coverage during the outbursts detected in IBIS.

The JEM-X data analysis shows no detections for \source\/, the mosaic image provides an upper limit of 9.6$\times$10$^{-13}$~erg~cm$^{-2}$~s$^{-1}$ in the 3-10 keV energy range (2$\sigma$ confidence level).

\subsection{Swift/XRT results }

\source\/ was observed with the XRT instrument $\sim$385 s after the BAT trigger on 2009 June 10 and during the following days.
The source showed a clear decline trend during this period with the brightest state observed in the first observation and the lowest
flux measured in the last pointing.
Unfortunately, no IBIS data simultaneous with the XRT observation were available but we can still
use these two observations to study the variability of the source.
Observations considered in this work were performed on Jun 10, 2009 07:38:00 and Jun 14, 2009 00:10:00 with an exposure time of 8221.5s and  2864.2 s, respectively.
The X-ray telescope  (XRT, Burrows et al. 2005) data reduction was performed using the XRTDAS standard data pipeline package ({\sc xrtpipeline}
v. 0.12.2), in order to produce screened event files. All data were extracted only in the Photon Counting 
(PC) mode (Hill et al. 2004), adopting the standard grade filtering (0--12 for PC mode), and analyzed with {\sc XIMAGE} version 4.4 in the 0.3--10 keV  energy band.
 We extracted the image from the first 630s of the first pointing where \source\/ was at the peak of the X-ray outburst with a 0.3-10 keV flux of
(2.4$\pm$0.1) $\times$10$^{-10}$~erg~cm$^{-2}$~s$^{-1}$.  During the second observation the flux had dramatically decreased being the source not detected by XRT: the 3$\sigma$ upper limit on the 0.3-10 keV flux was
1.4$\times$10$^{-12}$~erg~cm$^{-2}$~s$^{-1}$.
This is consistent with that measured by the JEM-X instrument.

\begin{table*}
\begin{center}
\caption{Summary of the Science Windows with detections 
of  \source\ }
\label{tab:det}
\centering
\begin{tabular}{cccc}
\hline
\hline
Obs ID         & IJD& exposure    &  ISGRI rate$^{(a)}$ \\
               & (=MJD--51544) &    (s)      &     (counts~s$^{-1}$)      \\

\hline
\multicolumn{4}{c}{First outburst {\scriptsize (2005-02-19 15:36:00.000 UTC)}: duration=$\sim$1 hour } \\

028700470010 & 1876.65   & 2576.0     & 3.8$\pm$ 0.6  \\

\hline
\multicolumn{4}{c}{Second outburst {\scriptsize (2009-01-28 23:40:33.880 UTC)}: duration=$\sim$3.5 hours }  \\

076800590010 &  3315.99 &  2270.2   &9.0$\pm$0.8  \\
076800600010 &  3316.03 &  2251.6 &7.2$\pm$0.8   \\
076800610010&   3316.07  & 2267.9  &4.1$\pm$0.9       \\
076800620010&   3316.11 &  2215.6 &6.4$\pm$0.9     \\

\hline
\multicolumn{4}{c}{Long time detection {\scriptsize (2005:241:10:19:12.000 UTC)}} \\

rev 351		& 2067.43&129430&0.61$\pm$0.08\\
\hline

\end{tabular}
\end{center}
\begin{small}
$^{(a)}$ in the 20-50 keV energy range \\
\end{small}
\end{table*}

Using the XRT fluxes we can compute a dynamic range (ratio between the outburst and quiescent fluxes) $>170$ for \source\/ which is
a value typical of known SFXTs.
This kind of X-ray transient activity is higher than that generally observed in other high mass X-ray binaries i.e. Be/X-ray transients. In fact
in this type of emitters their ratio between outburst and quiescence fluxes is generally below 20.

The XRT restricted position at the 99\% confidence level (RA=248.1579, DEC=-47.3937 with an error box of 2.7 arcsec) allows us to pinpoint the infrared counterpart of this source as the 2MASS object (2MASSJ16323791-4723409), in the
Spitzer source (GLIMPSEG336.7492+00.4223{\footnote{Galactic Legacy Infrared Mid-Plane Survey Extraordinaire (GLIMPSE) is available in the VizieR Service ({\em {http://vizier.u-strasbg.fr}})}}) and in the DENIS source (DENISJ163237.9-472341{\footnote{Deep Near-Infrared Survey of the Southern Sky,available in the VizieR Service: {\em {http://vizier.u-strasbg.fr}}}} ) with RA(J2000)=248.158 and DEC(J2000)=-47.395. This counterpart has magnitudes in the J, H, K bands of 14.631, 12.423 and 11.275 respectively.
Due to this absorption no optical counterpart was visible in UVOT images or listed in the USNO B1 catalogue.
A HMXB  nature would be consistent with the high level of absorption found in the soft X-rays (Grupe et al. 2009).

\section{Discussion}

Before the INTEGRAL era the known High Mass X-ray Binaries in our Galaxy were mostly Be X-ray Binary systems
and only a few supergiant systems were known as persistent but variable X-ray emitters.
The INTEGRAL satellite has changed our understanding of binary
systems, showing the existence of a new population of
compact objects orbiting
around a massive supergiant star, exhibits unusual properties, being either extremely absorbed,
or characterized by very short and intense flares.\\
In the scenarios proposed by Chaty et al. (2008),
 for  the classical super giant X-ray Binaries, the neutron star is orbiting at a few stellar radii only from the star.
The absorbed super giant X-ray Binaries, such as IGRJ16318-4848 (Filliatre and Chaty, 2004), are classical super giant X-ray Binaries hosting a neutron star constantly orbiting inside a cocoon made of dust and/or cold gas.

The SFXTs, consisting at the moment of $\sim$ 14 sources (Ducci et al. 2010), have peculiar behavior with rapid outbursts, faint
quiescent emission so far detected only in a few systems and hard X-ray spectra with outburst peak luminosity of $10^{36}$ erg/s.
With the increasing numbers discovered in the last 7 years it is now established that they can be divided in at least two different groups according with their $L_{min}/L_{max}$ value, duration and frequency of outburst: classical and intermediate SFXTs ( for review see Chaty  2010). In the classical SFXT, such as IGR J17544-2619 (Sidoli 2009, 2010),
the neutron star orbits on a large and eccentric orbit around the supergiant star, exhibits some recurrent and short transient X-ray flares and accretes from clumps of matter coming from the wind of the supergiant.
In Intermediate SFXT systems, as IGR J18483-0311 (Sguera et al. 2007, Romano et al. 2010), the neutron star orbits on a small and circular/eccentric orbit, and only when the neutron star is close to the supergiant star the accretion takes place and the source is active in the X-ray band.
Because it is passing through more diluted medium, the ratio between the outburst and quiescent luminosity
is higher for  classical SFXTs than for intermediate SFXTs.
In particular,  the typical ratio between quiescence
and outburst flux is less than 20 both in standard High Mass X-ray Binaries and in the absorbed systems, while is higher than 100 in
SFXTs.  \\
Although this scenario seems to describe quite well the characteristics currently seen in super giant X-ray Binaries, it very important to identify the nature of many more super giant X-ray Binaries to confirm this scenario and characterized their transient behavior to properly model their behavior.

 \source\/ was below the sensitivity of the IBIS instrument from 2003 to 2009 with a low quiescent emission ($<2.5\times10^{-12}~erg~s^{-1}~cm^{-2}$), then occasionally it underwent fast transient activity on different time scale: from
one hour to several hours and to several days. When the source becomes active its flux increased by a factor of more than $\sim$165 both in 0.3-10 keV and 20-50 keV bands.
 This is a typical behavior of the  Super Fast X-ray Transient sources (Grebenev  2009, Sguera et al. 2005, 2006,  Smith et al. 2006, Negueruela et al. 2006, Chaty 2010). Like many SFXT, \source\/ does not show any regularity in the outbursts, although the IBIS light curve discontinuity and the vicinity of the transient bright source 4U1630-472 prevent us from studying in detail the outburst periodicity.

Although the nature of IR counterparts of \source\/ have not yet been identified,
the transient and recurrent nature of the source, the spectral properties observed
during the outbursts lasting only for few tens of minutes/hours, the Galactic Plane location, the faint quiescent emission with a dynamic range $>$170 in the 0.3-10 keV and $>$165 in the 20-50 keV all point to \source\ being a member
of the class of Supergiant Fast X--ray Transients.

We note that the outburst durations for \source\/ are shorter than the typical outburst times ($\sim$ 1 day)
observed for other SFXTs. Assuming a distance of 10kpc, the luminosities are
$2.3\times10^{36}~erg~s^{-1}$, $4.0\times10^{36}~erg~s^{-1}$ and $5.3\times10^{35}~erg~s^{-1}$
for revolutions 287, 768 and 351, respectively. As reported in Grebenev and Sunyaev (2010) for the SFXT IGR~J18462-0223,
the short outburst duration could be a characteristic feature of the source \source\/ or an observational effect if the source is more distant than other SFXTs.

It is very important to study this class revealed by INTEGRAL
to define the dominant mechanism, explain the observed X- and hard X-ray behavior and confirm the scenario describing the normal SFXT and Intermediate ones and finally to understand the formation and evolution of the binary systems.



\acknowledgments
The authors acknowledge the ASI financial support
via ASI/INAF grants I/008/07/0/.

\end{document}